# Identifying Tipping Points in a Decision-Theoretic Model of Network Security

C. F. Larry Heimann, Alan Nochenson
Carnegie Mellon University
Information Systems
Pittsburgh, PA 15213
{profh, anochenson}@cmu.edu

*Abstract*—**Although system administrators are frequently urged to protect the machines in their network, the fact remains that the decision to protect is far from universal. To better understand this decision, we formulate a decision-theoretic model of a system administrator responsible for a network of size *n* against an attacker attempting to penetrate the network and infect the machines with a virus or similar exploit. By analyzing the model we are able to demonstrate the cost sensitivity of smaller networks as well as identify tipping points that can lead the administrator to switch away from the decision to protect.**

**Keywords-network security, decision theory, tipping points**

## I. INTRODUCTION

In an ideal world, system administrators would take any and every action necessary to protect the integrity of the machines under their jurisdiction. However, enhanced security always comes at a cost. Some of those costs come up front: there are costs to acquiring the appropriate security tools, costs with their installation and costs with maintaining these tools over time. At the same time, enhanced security can impose additional costs for those using the network. Security measures that make it difficult for intruders to penetrate a network can also decrease usability by making it somewhat more difficult for machines on the same network to connect and share information. Even though we do not live in a cost-free world, the conventional wisdom for system administrators is to take precautions to protect the network. For instance, Whitman and Mattord [7] argue that protecting one asset in a network may protect others as well. They cite the example of a firewall installed to defend against one type of attack which is also effective in preventing many other malicious attacks.

Despite this conventional wisdom, we know for a fact that the choice to protect computers on a network is far from universal. Gordon and Loeb [2] model information security from a purely economic standpoint and conclude that little to no security is economically necessary for extremely vulnerable and barely vulnerable systems. They show that limited resources may be better allocated towards systems that are more valuable (and hence harder to defend) or less valuable (because ensuring adequate security on highly sensitive data is very difficult). Kunreuther and Heal [6] examine a variety of interdependent security scenarios and also conclude that in certain scenarios protection is not the optimal strategy. Given these findings both for and against implementing protection, is it reasonable to try to identify possible tipping points that may lead a rational system administrator away from the decision to protect.

To identify potential tipping points, we have set up a decision-theoretic model building off of previous work done by Johnson, et al. [5]. In this paper, we look at the scenario where a network administrator is responsible for securing numerous computers and a faceless enemy is constantly attacking the network. The network is controlled by the administrator in a predictable way that can incur known costs with a known probability. The administrator only has information about the state of individual machines in the network, and how that state influences the chance of outside penetration. The attacker only knows the location of individual machines in the network, and a route to attempt to inject a virus into any given machine. His capabilities are restricted in that way in order to simplify the situation.

This paper is organized as follows: Section 2 discusses previous work in game theory and network security in order to put our model into context. Section 3 outlines the model and highlights key differences from previous models, while introducing the concept of a loss profile. Section 4 shows numerical and graphical analysis of optimal strategies as well as identifying tipping points for scenarios under various conditions. Section 5 discusses takeaway points and lessons learned and Section 6 outlines opportunities for further research on this subject.

## II. RELATED WORK

In their 2006 paper Alazzawe, et al. [1] discuss some of the issues associated with the use of intrusion and anomaly detection systems. Recognizing that a balance must be achieved between convenience and security, they then go on to formulate a game-theoretic model for determining if the security level of the entire system should be adjusted based on the readings of various sensors throughout the network. They conclude through simulation that a common intrusion detection system, Snort, performs better in adaptive (rule-based) mode. This suggests that game-theoretic formulations of tools perform better than their counterparts.

Grossklags, et al. [3] also use game theory to look at different security problems and classify these problems into five canonical security games: *total effort*, *weakest-link*, *best shot*, and *weakest target* (*with* and *without mitigation*). Each game shares the characteristic of relying on protection and insurance levels, and differs in how they are measured. For instance, in the *total effort* security game, the protection level of a system is a function of the mean protection level over each entity in the system. They then go on to find Nash equilibria strategies for each type of game. They end with discussion on intervention strategies applicable to all games in order to further network security within these scenarios. They acknowledge the role that limited information and uncertainty may play in these games, and set out to quantify these effects in further research.

Grossklags and Johnson [4] followed up on that by examining uncertainty in one of their security games, the *weakest link* game. This game is characterized by a tightly coupled interdependent network where the utility of a given player not only depends on his protection level, but that of his fellow users. They then go on to propose a revised utility model for this game, dependent upon the player's level of expertise. They also examine the role of information levels in the same game, and discuss the role information plays on the different types of players. One important conclusion they come to about information is that payoffs are no more than 18% better with complete versus incomplete information.

Johnson, et al. [5] continue looking into interdependent games, this time using a simpler model of a corporate LAN. They derive utility functions per game-player in homogeneous and heterogeneous situations based on a number of factors. They forgo the previous notions of insurance entirely, and look strictly at total protection from the outside world. Despite choosing to protect, a player still has the chance to be compromised from within the local network. They go on to discuss uncertainty in certain modeling parameters and use probability distributions to clear up some uncertainty, before computing Nash equilibria for the homogeneous and heterogeneous cases. They conclude that uncertainty in certain parameters does not significantly affect where the equilibria tend to lie.

## III. MODEL

We are looking at a scenario where $N$ users are closely connected through a local network. Each user maintains his own computer, which is connected to the outside world through an internet connection, which poses some risk of infection from external viruses. We call the probability of infection from an outside source $p$. Each user can eliminate the chance of outside infection by paying a fixed cost $c$ for a protection mechanism, e.g. virus protection software. For the simplicity's sake, we assume that buying the protection mechanism fully protects the user from outside threats. We make this assumption because by their nature, well-updated virus protection software should be close to completely secure. In order to keep everything on the same scale, we normalize $c$ such that $0 \leq c \leq 1$.

Each user $i$ also faces some probability $p_{ij}$ of infection from user $j$, for all $i \neq j$. In the homogenous situation, all $p_{ij}$ are equal. The probability of infection $p$ can also be thought of as $p_{ii}$, the probability of a user infecting himself from an outside source. In this paper, we only address only this homogenous situation, such that $q$ refers to any $p_{ij}$ where $i \neq j$. Previous analysis of a similar scenario [5] suggests that conclusions reached in the homogenous situations largely apply to the heterogeneous ones as well. Therefore, we omit that analysis for the sake of brevity.

Assuming homogeneity, let all users have the same chance of being infected from the outside, represented by $p$. We let $k \in [0, n]$ be the number of users who do not protect. Each of these $k$ users have the same probability $q$ of causing user $i$ to be infected through internal transmission, directly or indirectly. In this scenario there is no way a user can protect from internal infection, e.g. the virus protection software only monitors external connections. We expand upon a pre-established model [5] by adding terms $m_i$ and $n_i$. They are scaling parameters drawn independently from a probability distribution $D_i$, where each user can have their own distribution which is dictated by their loss profile.

All of these considerations lead to the following model. The expected utility for user $i$ is given by:

$$U_i = \begin{cases} -c + m_i * (1-q)^k & \text{, if user } i \text{ protects} \\ n_i * (1-p)(1-q)^k & \text{, if user } i \text{ does not protect} \end{cases}$$

Before proceeding further, it is necessary to formally introduce the notion of a loss profile, and to show its influence on the above model. A loss profile $L_i(x)$ is a discrete probability distribution over potential loss for player $i$ such that $P(l = x) = L_i(x)$. The terms $m_i$ and $n_i$ in the above equation are drawn from a player's individual loss profile $L_i$. In most cases $m_i \neq n_i$, since they are drawn independently from player $i$'s loss profile. In the context of the model, if a user protects, he will face a possible loss from internal infection, the magnitude of which will be dictated by $m_i$. If he does not protect, he will still face a possible loss from both internal and external infections, scaled by $n_i$.

For this paper, we restrict our analysis an 'all-or-nothing' loss profile where a virus either causes total damage ($x = 1$) or no damage ($x = 0$). We assume that every user has this same profile. This profile would be represented by a distribution where each outcome is equally likely, or $L_i(x) = 0.5 \ \forall x \in \{0,1\}$. One situation where this profile is applicable is the case where a server is responsible for containing confidential information. If a virus infects this server, confidential data is leaked, which can be regarded as a total loss. There is also the possibility that despite breaching the wall of the intranet, a virus does not do any damage. This would be possible if the virus could not handle user $i$'s computer, e.g. the virus was not written for the correct operating system.

An important distinction that we make from previous work is the concept of a single decision-maker, e.g. a system administrator for the corporate network. She is responsible for mandating protection on a per-user basis. In terms of our model, the parameters $c$, $p$, and $q$ are known quantities that are uniform across all users, and $k$ is chosen by the decision-maker in order to maximize overall expected system utility. This is particularly noteworthy because a single decision-maker-scenario means that all expected utilities per user are calculated at the same point in time, having no concrete information about the other users, other than the probabilities of $c$, $p$, and $q$, and the distributions which $m_i$ and $n_i$ are drawn from for each user.

## IV.    ANALYSIS

### A.  *Averaging*

In order to analyze the model in a meaningful way, we need to understand the users not as a collection of individuals; we partition the users into two distinct sets. We place users into one of the sets based on their choice to protect or not (defect). Using the same notion as before, the mean utility in a system is given by:

$$U_{all} = U_{protect} + U_{defect}$$

$$= \sum_{i=1}^{N-k} U_i + \sum_{i=1}^{k} U_i$$

$$= \sum_{i=1}^{N-k} -c + m_{avg}(1-q)^k + \sum_{i=1}^{k} n_{avg}(1-p)(1-q)^k$$

$$U_{all} = (N-k)\left(-c + m_{avg}(1-q)^k\right) + k * n_{avg}(1-p)(1-q)^k$$

The decision-maker is trying to maximize the value of $U_{all}$ by choosing the optimal $k \in [0, N]$. In other words, he is trying to maximize the overall utility by choosing how many users in his network should not protect (or should protect. They problems are equivalent). Previous research [5] suggests that optimal utility will not be found at some non-boundary case. This leads us to restrict our analysis to situations where $k = 0 \ or \ k = N$, everyone protects or everyone defects, respectively. When $k = 0$ and everyone protects, overall utility reduces to:

$$U_{all|protect} = (N-0)\left(-c + m_{avg}(1-q)^0\right) + 0 * n_{avg}(1-p)(1-q)^0$$

$$U_{all|protect} = N * \left(-c + m_{avg}\right)$$

When $k = N$ and everyone defects, overall utility reduces to:

$$U_{all|defect} = (N-N)\left(-c + m_{avg}(1-q)^N\right) + N * n_{avg}(1-p)(1-q)^{N^n}$$

$$U_{all|defect} = N * n_{avg}(1-p)(1-q)^N$$

A visualization of the all-defect strategy is shown in Figure 1. A visualization of the all-protect strategy is shown in Figure 2. Note that the graph of utility for the all-protect strategy has numerous lines for various values of c (protection costs); while the all-defect graph does not have these lines. This is because in the all-defect strategy, utility is not a function of the protection cost, because not even a single user protects.

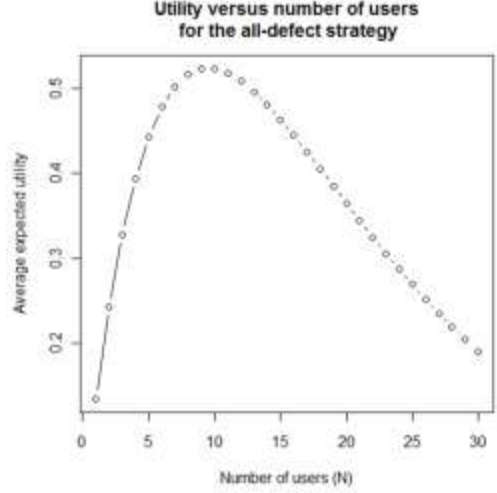

Figure 1

*B. Simplification for the all-or-nothing loss profile*

In this paper, we restrict the distribution we draw the scaling parameters from. The parameters $m_i$ and $n_i$ are drawn from the set {0,1} with equal probabilities. Therefore, for the purpose of this analysis, we can replace $m_{avg}$ and $n_{avg}$ with 0.5. This substitution yields:

$$U_{all|protect} = N(0.5 - c)$$

$$U_{all|defect} = 0.5N(1-p)(1-q)^N$$

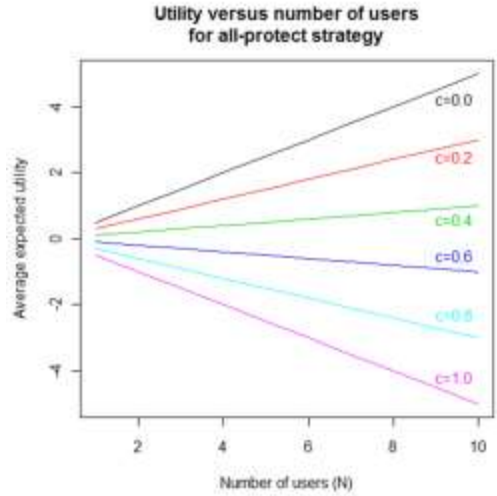

Figure 2

## C. Tipping Points

We have now derived formulas for Utility given the all-protect and all-defect strategies. We now need to figure out when one is preferable over the other. We define a *tipping point* as a situation where the overall system utility for the all-protect scenario equals that of the all-defect scenario, and thus a switch in the optimal strategy will occur. In other words:

$$U_{all|protect} = U_{all|defect}$$

$$N(0.5 - c) = 0.5N(1 - p)(1 - q)^N$$

$$0.5 - c = 0.5(1 - p)(1 - q)^N$$

$$1 - 2c = (1 - p)(1 - q)^N$$

$$\frac{1 - 2c}{1 - p} = (1 - q)^N$$

$$\log_{1-q} \frac{1 - 2c}{1 - p} = N$$

In Figure 3, we plot expected utility as a function of the number of users, *N*, for $c = 0.4$. We show the all-defect and all-protect cases as separate lines, which can be seen as taken from Figures 1 and 2, respectively. The intersection of the lines is the tipping point for this case, where the all-protect strategy becomes preferable to the all-defect strategy (which happens around 4 users, with an expected utility of around 0.4).

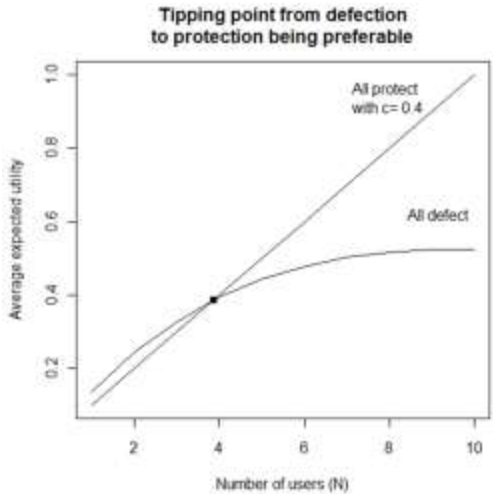

Figure 3

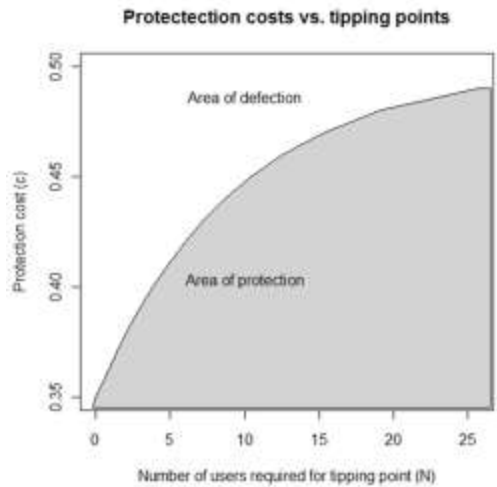

Figure 4

It is noteworthy that in Figure 1, the utility in the all-defect strategy approaches 0 as N approaches infinity. Therefore, it is not always possible to find a number of users $N$ given a protection cost $c$, such that optimal strategy will change. In some scenarios, it is always better to protect, or always better to defect, regardless of how many users there are. Figure 4 illustrates this, showing the minimum number of users $N$ needed for a tipping point to exist for a given value of $c$. In this example, where $p = 0.5$ and $q = 0.1$, there are no tipping points when the protection cost $c \notin (0.35, 0.5)$. This is a logical extension from the equation above. We say that no tipping point exists when $N$ is non-positive or infinity.

## V. Conclusions

In the all-protect scenario, the average expected utility of a system is scaled by the number of users in the network. In other words, an all-protect scenario that has a constantly increasing utility (i.e., c=0.2) will continue to increase as the number of users increase. This can be used to infer relationships between utilities for networks with different numbers of users. In general, average expected utility for an all-protect scenario is a monotonic function, which is non-increasing if $c \geq 0.5$ and increasing if $c < 0.5$. Utility for the all-defect scenario has a much smaller range, and approaches 0 as the number of users gets larger. This motivates the concept of cost sensitivity based on the number of users in a network.

As is illustrated in figure 4, smaller networks are more vulnerable to fluctuating protection costs than large networks. To better understand this, all area above the line in figure 4 can be thought of as the *area of defection*, named for the dominant strategy in that area. All area below then line is the *area of protection*, where protection is the dominant strategy. Take for example the case of $N = 5$ *and* $c \approx 0.4$. Defection is the optimal strategy for close to 60% of possible protection cost values.

This contrasts with a larger network, where the line has a horizontal asymptote at 0.5, and neither strategy is dominant for a uniformly distributed protection cost.

We derive here two main results, one "good," and the other "bad." The "good" news is for all areas outside of the graph, $c \notin (0.35, 0.5)$, there exists a dominant strategy. This makes the decision-maker's job very easy. For 85% of the values of c, he is given a logical way to act regardless of the number of users in his network. For the other 15%, the dominant strategy is based on the number of users, and is illustrated above. The "bad" news is that the strategy which is dominant in more cases is the defection strategy. All $c > 0.5$ force defection regardless of the number of users involved, which is half of the sample space. The contested area $0.35 < c < 0.5$ tends toward defection almost half of the time, leading to an overall estimated 60% of time that the defection strategy is dominant.

The frequency at which administrators choose not to protect in our model is consistent with other work like Kunreuther and Heal [6], even if it flies in the face of conventional wisdom. Of course, conventional wisdom may also factor in certain psychological benefits (an administrator sleeps better at night knowing the network is secure) as well as political one (easier to shift blame to the tools in the case of a breach) -- factors we have not taken into consideration in our model. But the fact is a significant portion of administrators in the real world do choose to forego protection, supporting our finding that protection is not always the optimal choice for a given network. Knowing the characteristics of the network, our model would allow administrators to better identify the tipping points where the decision to protect or not changes direction.

VI.    FUTURE RESEARCH

In this model we introduced the concept of a loss profile, which opens up many new avenues for future research. It is natural to expect that variations in loss profiles may lead to different outcomes. Moreover, in this paper we have treated the population as homogeneous, with all the participants sharing the same loss profile. In a more heterogeneous population where participants had different loss profiles, the results could be influenced by the percent of people who fell into profile strata.

In a related manner, our paper has a single decision-maker, i.e., the system administrator, who could act on behalf of the entire network. If we allowed for individuals within the network to act on their own, we introduce a whole new level of complexity. The strategy space for a player would not only contain the decision to protect or defect, but when to do so. Under this new model, the game would become a sequential one where individuals have to decide not only if to protect, but when to protect. By delaying the decision to protect, a player gains information about other actors and their choices, but also risks infection from the attacker.  This also opens up the possibility of negotiating binding or non-binding agreements before acting. All of these are promising areas that could yield some interesting results.